\documentclass[3p,times]{elsarticle}

\usepackage{no-ecrc}

\volume{00}

\firstpage{1}

\journalname{Computer Physics Communications}

\runauth{S. Frijters et al.}

\jid{cpc}

\jnltitlelogo{Comput. Phys. Commun.}

\CopyrightLine{2014}{Published by Elsevier Ltd.}

\usepackage{amssymb}

\usepackage[figuresright]{rotating}

\usepackage{subcaption}

\usepackage{color}

\renewcommand{\vec}[1]{\mathbf{#1}}

\begin{document}

\begin{frontmatter}



\dochead{}

\title{Parallelised Hoshen-Kopelman algorithm for lattice-Boltzmann simulations}


\author[label1]{S. Frijters}
\author[label2]{T. Kr\"uger}
\author[label1,label3]{J. Harting}

\address[label1]{Department of Applied Physics, Eindhoven University of Technology, Den Dolech 2, 5600MB Eindhoven, The Netherlands}
\address[label2]{Institute for Materials and Processes, School of Engineering, University of Edinburgh, Mayfield Road, Edinburgh EH9 3JL, Scotland, UK}
\address[label3]{Faculty of Science and Technology, Mesa+ Institute, University of Twente, 7500 AE Enschede, The Netherlands}

\ead{j.harting@tue.nl}

\begin{abstract}
We discuss two topics that we have encountered in our lattice-Boltzmann simulations of complex fluids: the sizes of droplets in particle-stabilised emulsions and deformable particles in fluid flow. The common factor in these seemingly disparate subjects is that both represent an opportunity for a novel application of the Hoshen-Kopelman algorithm. This algorithm is based on detecting connected clusters on a lattice and labelling the involved lattice sites such that all sites that are connected share the same label. The assumption of the presence of a lattice makes it a convenient algorithm to use in combination with lattice-Boltzmann simulations. In order to apply this algorithm on the fly during massively parallel simulations, it needs to be parallelised as well. We present our parallel implementation, which is tailored to a common parallelisation scheme for the lattice-Boltzmann method, and compare it to previous work. We then briefly discuss some examples of results obtained using this procedure.
\end{abstract}


\end{frontmatter}



\section{Introduction}

Many subtopics of the field of soft matter are of interest not only academically, but also because of direct applications. In many cases, not all properties of the relevant systems are easily accessible in experiments. This is where computer simulations can supply valuable insights. In this work, we discuss two such applications that have been studied using our lattice-Boltzmann (LB) based simulation code: in recent decades, many branches of industry have started using nanoparticles to stabilise emulsions for various purposes (\emph{e.g.}~cosmetics, improved low-fat food products, drug delivery), while the behaviour of deformable particles in fluid flow is of interest to, for example, medical research. These two topics have led to new applications of the Hoshen-Kopelman (HK) algorithm~\cite{ bib:hoshen-kopelman:1976}: obtaining statistics on the sizes of individual droplets in (particle-stabilised) emulsions and determining the interior of deformable particles in fluid flow.

The work of Hoshen and Kopelman has been described by the authors as a novel
``cluster multiple labeling technique''. The algorithm was developed to apply
to quantities on two- or three-dimensional lattices and is based on detecting
connected clusters on a lattice and labelling the involved lattice sites such
that all sites that are connected share the same label. It was originally
intended as an aid to the study of percolation of porous media but has since
been used in a wide variety of fields and for a wide range of applications,
such as nuclear fuel rod processing, catalysis, development of bone structures
and the consistency of fried foods~\cite{ bib:hoshen-berry-minser:1997}. The
assumption of the presence of a lattice makes it easy to adapt the algorithm
for use in combination with LB-based simulations. In order to apply this
algorithm on the fly during massively parallel simulations, it needs to be
parallelised as well. With the advent of parallel computations, methods have
been introduced to achieve this, however, none of them makes effective use of
the parallelisation structure of the LB method~\cite{bib:burkitt-heermann:1989,
bib:flanigan-tamayo:1995, bib:berry-constantin-vanderzanden:1997,bib:teuler-gimel:2000}. Therefore,
we present our implementation here and compare it to previous work.

This paper is structured as follows. First, the serial HK algorithm is explained in some detail. Then, the parallelisation scheme used in our simulation code is introduced and a set of benchmark results and a qualitative comparison to other schemes are supplied. We touch upon a few salient points of the LB method and some of its extensions, which are used to simulate the following two systems: particle-stabilised emulsions and capsules/vesicles in fluid flow. Examples of these systems are then discussed, highlighting new applications of the HK algorithm.

\section{Serial Hoshen-Kopelman algorithm}

\begin{figure}[!htbf]
  \centering
  \begin{subfigure}[b]{0.24\textwidth}
    \centering
    \includegraphics[scale=0.75]{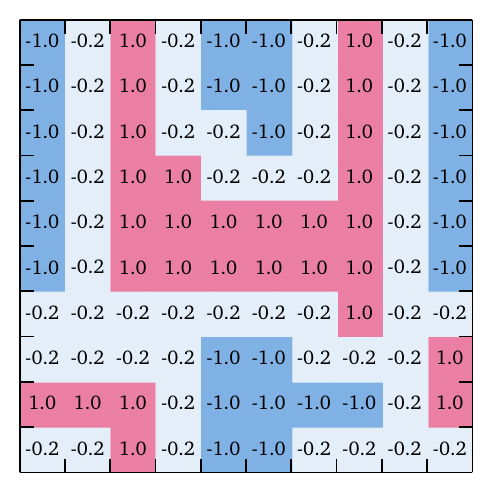}
    \caption{}
    \label{fig:hk-fields-1}
  \end{subfigure}
  \begin{subfigure}[b]{0.24\textwidth}
    \centering
    \includegraphics[scale=0.75]{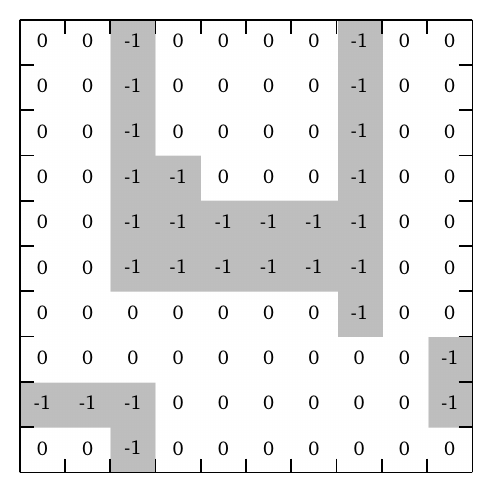}
    \caption{}
    \label{fig:hk-fields-2}
  \end{subfigure}
  \begin{subfigure}[b]{0.24\textwidth}
    \centering
    \includegraphics[scale=0.75]{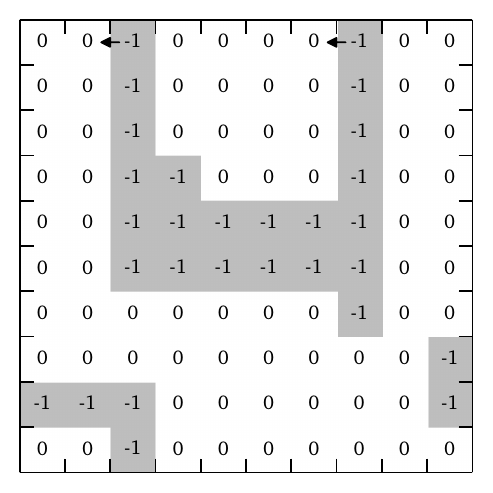}
    \caption{}
    \label{fig:hk-fields-3}
  \end{subfigure}
  \begin{subfigure}[b]{0.24\textwidth}
    \centering
    \includegraphics[scale=0.75]{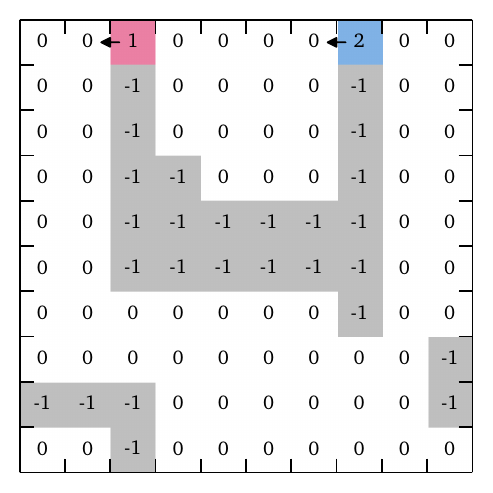}
    \caption{}
    \label{fig:hk-fields-4}
  \end{subfigure}
  ~\\
  \begin{subfigure}[b]{0.24\textwidth}
    \centering
    \includegraphics[scale=0.75]{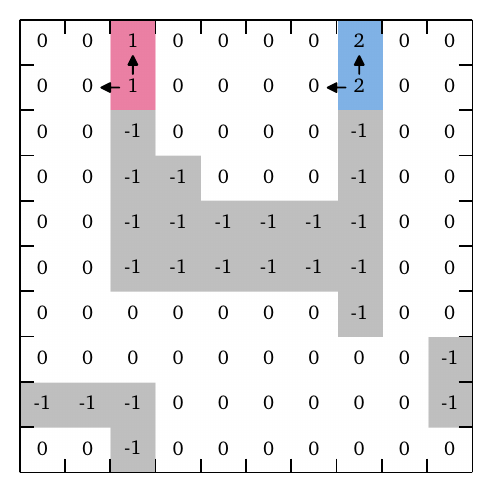}
    \caption{}
    \label{fig:hk-fields-5}
  \end{subfigure}
  \begin{subfigure}[b]{0.24\textwidth}
    \centering
    \includegraphics[scale=0.75]{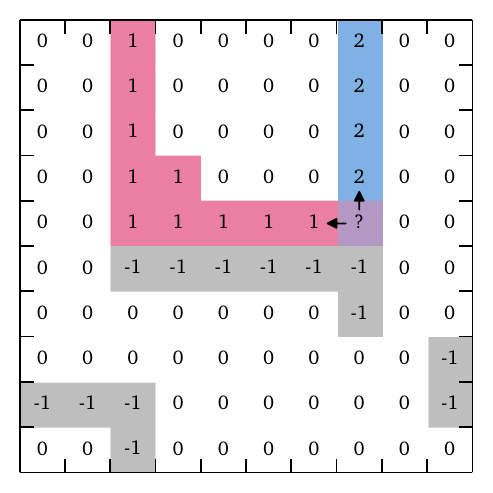}
    \caption{}
    \label{fig:hk-fields-6}
  \end{subfigure}
  \begin{subfigure}[b]{0.24\textwidth}
    \centering
    \includegraphics[scale=0.75]{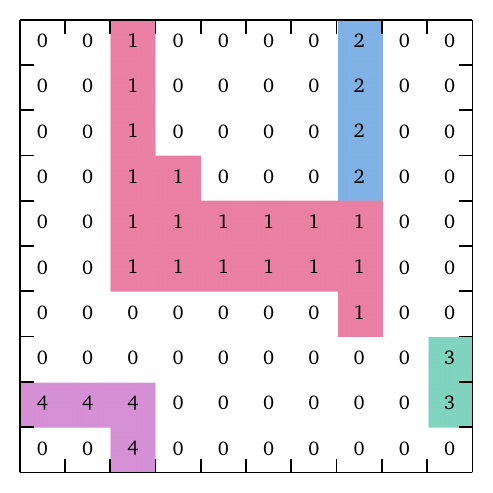}
    \caption{}
    \label{fig:hk-fields-7}
  \end{subfigure}
  \begin{subfigure}[b]{0.24\textwidth}
    \centering
    \includegraphics[scale=0.75]{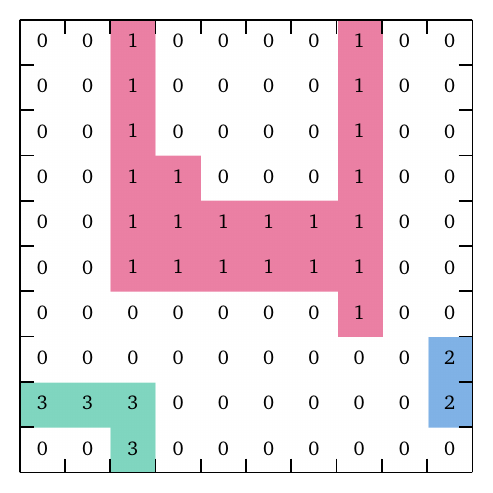}
    \caption{}
    \label{fig:hk-fields-8}
  \end{subfigure}

  \caption{The Hoshen-Kopelman algorithm needs a condition to determine whether or not a lattice site is to be considered part of a cluster. Without loss of generality we assume that for some order parameter $\varphi(\vec{x})$ the condition is $\varphi(\vec{x}) > 0$. An example of such a field in 2D is shown in (\subref{fig:hk-fields-1}). As an initial condition for the algorithm, a value of $l(\vec{x}) = 0$ is assigned to sites that are not part of any cluster and a value of $l(\vec{x}) = -1$ to those that are. The results of this initialisation are shown in (\subref{fig:hk-fields-2}). After initialisation, the HK algorithm runs over the lattice and applies a set of rules. In this example, we pass over the lattice row-first and generally consider sites to the left and above a site to be neighbours (\subref{fig:hk-fields-3}). In the top row, no sites exist above. Application of the first rule (\subref{fig:hk-fields-4}): if none of the neighbours has a value of $l_{\mathrm{nb}} > 0$, assign $l(\vec{x}) = l_{\max} + 1$, where $l_{\max}$ is the largest label currently assigned to any lattice site. Application of the second rule (\subref{fig:hk-fields-5}): if exactly one of the neighbours has a value of $l_{\mathrm{nb}} > 0$, assign $l(\vec{x}) = l_{\mathrm{nb}}$. Application of the third rule (\subref{fig:hk-fields-6}): if more than one neighbour has a value of $l_{\mathrm{nb},i} > 0$, assign $l(\vec{x}) = \min(l_{\mathrm{nb},i})$. We also store the equivalence relations $l_{\mathrm{nb},i} \equiv l_{\mathrm{nb},j}$ with $i$ and $j$ running over all neighbours. In this case we set the value of the site to $1$ and store $1 \equiv 2$ for later use. This prodecure continues until all sites have been visited (\subref{fig:hk-fields-7}). The clearly connected cluster at the top now has some sites with $l = 1$ and some with $l = 2$, which is not desirable. However, this is solved in the last step of the algorithm: the previously stored equivalence relations (in this example: $1 \equiv 2$) are compactified and applied, resulting in correctly labelled clusters (\subref{fig:hk-fields-8}).
}
  \label{fig:hk-fields}
\end{figure}

The HK algorithm consists of an initialisation step, three rules that are applied to all sites of the lattice and a relabelling step at the end. Before initialisation can take place, one needs to choose a condition to decide if a particular site is to be considered part of a cluster or of the medium. With a view towards the applications to be discussed later, we introduce an order parameter $\varphi(\vec{x})$, which is a real number defined on each lattice site. The required condition is now chosen to be based on the local value of this order parameter: sites $\vec{x}$ for which $\varphi(\vec{x}) > 0$ are part of a cluster, while $\vec{x}$ for which $\varphi(\vec{x}) \le 0$ are not part of any cluster. As an initial condition for the algorithm, a label value of $l(\vec{x}) = 0$ is assigned to sites that are not part of any cluster and a value of $l(\vec{x}) = -1$ is temporarily assigned to those that are. This process is illustrated in Fig.~\ref{fig:hk-fields-1} and Fig.~\ref{fig:hk-fields-2}; in this example, we show a finite bounded two-dimensional lattice.

After this initialisation procedure the algorithm runs over the lattice, and at every visited lattice site $\vec{x}$, if the current label is $l(\vec{x}) = -1$, the following three rules are applied:
\begin{enumerate}
\item If none of the neighbours has a value of $l_{\mathrm{nb}} > 0$, assign $l(\vec{x}) = l_{\max} + 1$, where $l_{\max}$ is the largest label currently assigned to any lattice site.
\item If exactly one of the neighbours has a value of $l_{\mathrm{nb}} > 0$, assign $l(\vec{x}) = l_{\mathrm{nb}}$.
\item If more than one neighbour has a value of $l_{\mathrm{nb},i} > 0$, assign $l(\vec{x}) = \min(l_{\mathrm{nb},i})$. At this point, parts of the same cluster have been assigned different labels, which is not desirable. To correct this in the final step of the algorithm, this relation is stored as a set of equivalence relations $l_{\mathrm{nb},i} \equiv l_{\mathrm{nb},j}$ with $i$ and $j$ running over all neighbours.
\end{enumerate}
In this context, sites that are diagonally adjacent to each other are not considered to be neighbours, and one only looks ``backwards'' on the lattice (\emph{i.e.}~left and up in this example). The application of these three rules is illustrated in Fig.~\ref{fig:hk-fields-3} through Fig.~\ref{fig:hk-fields-6}.

After application of these three rules to all lattice sites, we are left with only positive values of $l(\vec{x})$ and generally a non-empty set of equivalence relations. As a final step, the lattice is looped over one last time, and at every lattice site the equivalence relations are applied in a way such that every connected cluster has the smallest possible number. One can also choose to first reduce the assigned labels in the equivalence relations in such a way that all integers $1, \ldots, n$ are used after applying them, where $n$ is the number of clusters on the lattice. This compactification is normally not needed, as a label is just a label, but as will be described below, it simplifies the parallelised version of the algorithm. This process and the resulting final state are shown in Fig.~\ref{fig:hk-fields-7} and Fig.~\ref{fig:hk-fields-8}.

\section{Parallelised Hoshen-Kopelman algorithm}
\label{ssec:emulsions-phk}

\begin{figure}[!htbf]
  \centering
  \begin{subfigure}[c]{0.3\textwidth}
    \centering
    \includegraphics[scale=0.75]{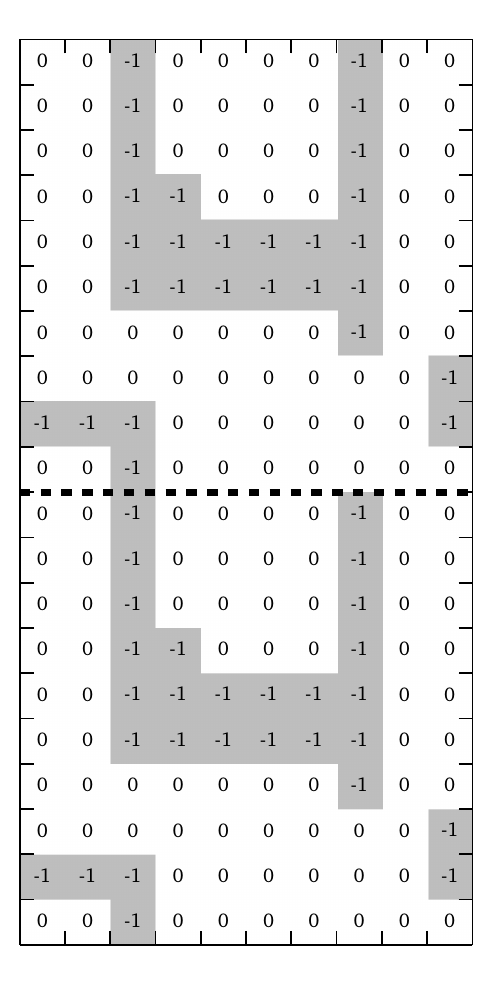}
    \caption{}
    \label{fig:hk-par-1}
  \end{subfigure}
  \begin{subfigure}[c]{0.3\textwidth}
    \centering
    \includegraphics[scale=0.75]{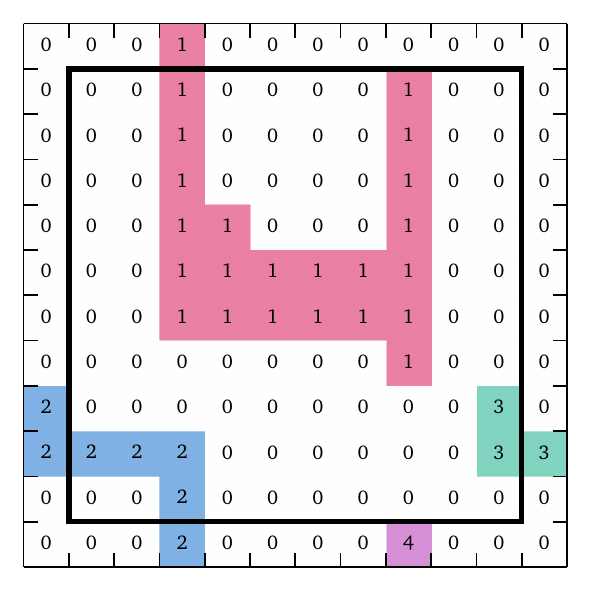}
    \includegraphics[scale=0.75]{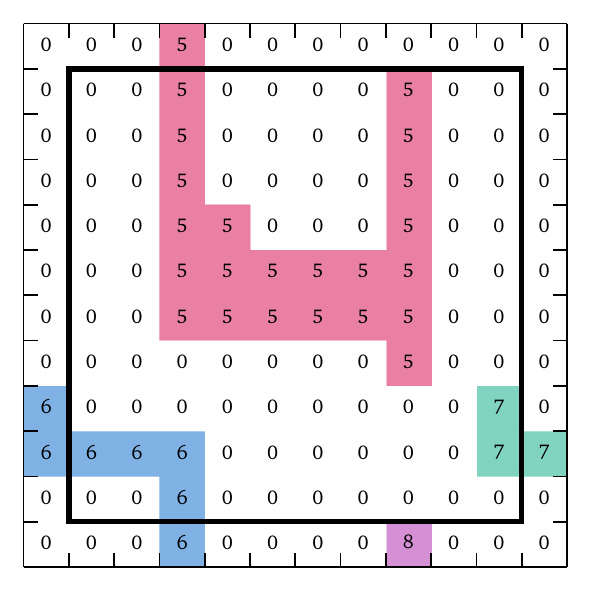}\\
    \caption{}
    \label{fig:hk-par-2}
  \end{subfigure}
  \begin{subfigure}[c]{0.3\textwidth}
    \centering
    \includegraphics[scale=0.75]{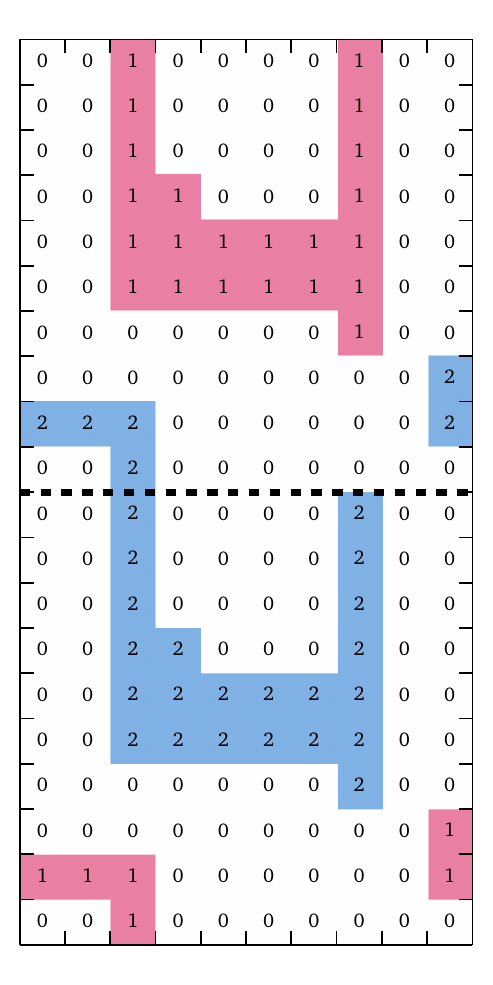}
    \caption{}
    \label{fig:hk-par-3}
  \end{subfigure}
  \caption{To show the process of the parallel HK algorithm we use the system (\subref{fig:hk-par-1}) as an example. The dashed line indicates the division of the system over two parallel processes, and the outer edge now represents periodic boundary conditions. Each process includes a ``halo'' region and follows the serial HK algorithm (\subref{fig:hk-par-2}). They assign a range of labels, offset by a value such that the values that various processes can assign do not overlap. In this case, the values of process 2 (bottom) are offset by $4$, being the largest number of clusters found by any one process. As with the serial version of the algorithm, equivalence classes are used, but in this case they are also based on the labels of halo sites and their matching physical counterparts. In this way, the finally assigned labels are globally correct (\subref{fig:hk-par-3}).}
  \label{fig:hk-par}
\end{figure}

As the systems of interest in our research are large, calculating the cluster indices during the simulation, when each domain is still assigned to its own CPU, rather than during post-processing, is very desirable. To that effect, the HK algorithm has been extended to work in a parallel fashion. Methods to achieve this have been succesfully used in previous works~\cite{bib:burkitt-heermann:1989, bib:flanigan-tamayo:1995, bib:berry-constantin-vanderzanden:1997,bib:teuler-gimel:2000}, but to the best of the authors' knowledge, no new advances have been made recently. Previous solutions have opted to increase the complexity of the algorithm to gain higher performance; however, we have found that on current hardware, a more straightforward implementation will suffice in many cases. In particular, we use an implementation that meshes very well with a common parallelisation structure of the lattice-Boltzmann method.

Consider the example system shown in Fig.~\ref{fig:hk-par-1} --- it is twice the size of the previous example and we want to divide the workload over two processes, as indicated by the dashed line. In addition, the total system is now subject to periodic boundaries in both directions, a situation which mimics the systems that will be described later in this work (those will be fully three-dimensional with periodic boundary conditions in at least some directions). To facilitate the communication between the processes, the calculation domain of each process is extended by a layer one lattice site wide on each side, and this extra layer corresponds to the outermost layer of sites of the neighbouring process (not counting the additional layer). These extra sites are referred to as ``halo'' sites (also known as ``ghost cells''~\cite{ bib:kjolstad-snir:2010}), while the original ones are called ``physical'' sites. This extension and its nomenclature follow the use of halos in the LB method (and similarly structured algorithms).

As in the case of serial HK, each process is assumed to hold the value of the order parameter on its extended lattice. The HK algorithm is then executed by each process on its local (extended) domain, and the maximum cluster indices are reduced as much as possible. Then, each process reports to each other process the value of their highest assigned cluster id $n_{\mathrm{max}}$ (which corresponds to the number of clusters found in that domain). The highest global id is then used by each process to apply a transformation to each local cluster id:
\begin{equation}
  l_{\mathrm{gl}} = l_{\mathrm{loc}} + \mathrm{rank} \cdot n_{\mathrm{max, gl}}
  \mbox{,}
\end{equation}
where $l_{\mathrm{gl}}$ is now guaranteed to be a globally unique id, and $\mathrm{rank}$ is the rank of the process on which the site resides. The result of this procedure can be seen in Fig.~\ref{fig:hk-par-2}. Each process then communicates the value of the cluster id on sites in the halo to its neighbours; the data corresponds to the data in the physical region of the neighbour, as described above. This correspondence is expressed by creating additional equivalence classes, by comparing the cluster values calculated on the physical sites to the values received by the halo exchange. In this example, a number of such equivalences is generated: $1 \equiv 6$, $1 \equiv 8$, $2 \equiv 5$, $4 \equiv 5$ (top-bottom), and $2 \equiv 3$ and $6 \equiv 7$ (left-right). All equivalence classes are then communicated to the root process and combined in such a way that the values of the labels are minimised. These are then scattered to the relevant processes (each process only needs equivalence classes for global indices $\mathrm{rank} \cdot n_{\mathrm{max, gl}} \le l_{\mathrm{gl}} < ( \mathrm{rank} + 1 ) \cdot n_{\mathrm{max, gl}}$) and applied. After this step the algorithm is complete and, in this example, we are left with only two clusters, \emph{cf.}~Fig.~\ref{fig:hk-par-3}.

\begin{figure}[!htbf]
  \centering
  \includegraphics[scale=0.75]{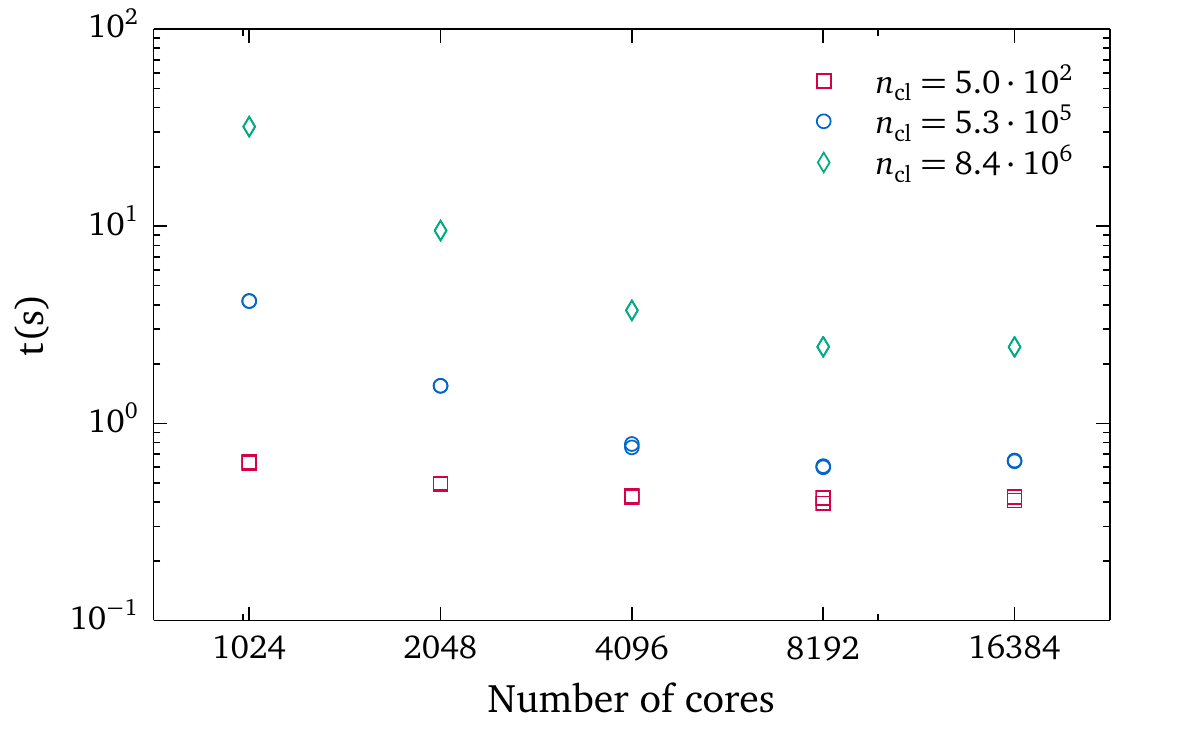}
  \caption{Strong scaling of the parallelised Hoshen-Kopelman algorithm for systems of volume $V = 1024^3$ containing $n_{\mathrm{cl}}$ clusters, as performed on on JUQUEEN, the Blue Gene/Q supercomputer at the J{\"u}lich Supercomputing Centre. The work to be done depends on the number of clusters in the system, and, as some work is done by the root process only, the scaling is not linear for large numbers of processors. However, for applications such as the ones described in the following sections, the absolute times required to perform the algorithm are negligible compared to execution times of the calculations related to the physics simulation.}
  \label{fig:hk-timings-strong}
\end{figure}

We have performed benchmarks on JUQUEEN, the Blue Gene/Q supercomputer at the J{\"u}lich Supercomputing Centre, which employs 1.6 GHz IBM PowerPC A2 cores, and varied the number of used cores between 1024 and 16384. The benchmarks are based on a synthetic test case of a system consisting of $1024^3$ lattice sites, filled with rectangular domains touching only at the corners (such that they are not connected in the sense of the HK algorithm) and alternating between $\varphi(\vec{x}) > 0$ and $\varphi(\vec{x}) \le 0$. This results in a deterministic and regular system, which remains invariant when the number of processes it is distributed over changes. The strong scaling for such systems with different cluster sizes (and thus different numbers of clusters $n_{\mathrm{cl}}$) is shown in Fig.~\ref{fig:hk-timings-strong}. We have observed that up to $O(10^5)$ clusters the algorithm takes almost constant time. For larger numbers of clusters, a power law has been observed, reducing the efficiency of the implementation but not rendering it prohibitively expensive until $O(10^8)$ clusters and beyond. Keep in mind, however, that for a system of fixed size $V$ a theoretical maximum number of clusters exists: $n_{\mathrm{cl, max}} = V/2$ (this corresponds to a system of clusters of size $1$, positioned in an alternating pattern in all dimensions). In this case, it means that only the last decade of possible numbers of clusters is strongly affected. To put this further into perspective: in our use cases described below, $n_{\mathrm{cl}}$ will not exceed $10^6$ and will more usually be below $10^4$. It is expected that the observed power law behaviour is caused by the structure of the algorithm, but only by the underlying implementation. As such, this bottleneck could be removed in the future, for applications that require the use of so many small clusters. Similarly, the memory footprint of the algorithm can be reduced by optimising the data structures in use.

It has been found that efficient parallelisation according to the method described above needs some careful tuning. In particular, when using a naive array-based implementation for equivalence classes, and using the root process to combine all global equivalence relations, it is mandatory to re-order the cluster ids such that they are minimised, as the length of the array depends linearly on both the number of processes and the number of clusters per process. This reduces the sizes of the required arrays and reduces the volume of communication significantly. Local equivalence relations should be applied before parallelisation, for the same reason. Equivalence classes can be represented by a singly linked list, such that a traversal of the list, starting from some arbitrary cluster id, ends at the lowest value of all of its combined equivalence relations. Adding a new equivalence relation that connects two clusters can then be resolved by traversing the list starting at both cluster ids, determining the smallest final value, and updating the list ending in the larger value. The linked list can then be traversed one more time for each cluster id to reduce it to its minimal equivalent id. The linked list implementation also serves to reduce the variance expected when the shape of clusters changes, \emph{i.e.}~clusters spanning many processes, creating many equivalence relations. We conclude that this algorithm is easily implemented, very well suited to the structure of the LB method, and efficient as long as the number of clusters is not excessive (relative to the size of the lattice).

The method presented here is similar to the method proposed by Burkitt and
Heermann~\cite{ bib:burkitt-heermann:1989}, who prefer to avoid a potentially
expensive gather and scatter operation as well as computational imbalance at
the cost of a more complicated communication structure and algorithm (this
complexity would increase even more when considering that their original work
is done in 2D). We have found that for our purposes, the communication is not a
bottleneck and even a single process is very efficient at processing the
relations between all partial clusters. As such, ease of implementation is
preferred. Flanigan and Tamayo~\cite{ bib:flanigan-tamayo:1995} have proposed a
relaxation scheme, in which processes communicate repeatedly with their
neighbours, propagating the lowest cluster id of a cluster spanning multiple
processes until a termination condition (no process has changed any of its
cluster ids) is reached. The efficiency of this method is expected to decrease
as the spatial decomposition in all three dimensions is no longer uniform, as
changes will have to propagate also in the direction with the most processes.
Teuler and Gimel~\cite{bib:teuler-gimel:2000} have presented a scheme based on
cyclic reduction of the cluster labels. The
method is tailored for efficiency on distributed memory machines, but requires
the domain decomposition to be one-dimensional (slabs). This makes the
algorithm not generally suitable for application in the context of simulations
utilizing a fully three-dimensional cubic domain decomposition scheme.
Our current implementation also resembles the work of Berry et al.~\cite{
bib:berry-constantin-vanderzanden:1997}, who perform the final merge
calculation on the root process as well, but use a different parallel
communication structure, based on neighbouring sites rather than overlapping
(halo) sites.

\section{The lattice-Boltzmann method}

Our simulations are focused on various complex fluids, which are fluid mixtures which may include --- among others --- multiple fluid components, solid particles, or deformable particles. To resolve the hydrodynamics our method of choice is the lattice-Boltzmann (LB) method, which is an alternative to traditional Navier-Stokes solvers. The method was chosen for a number of reasons: it is well-established in the literature~\cite{ bib:chen-doolen:1998, bib:succi:2001, bib:sukop-thorne:2007, bib:benzi-chibbaro-succi:2009}, and extensions beyond single-component fluid flows have been developed to allow for multiple fluids and their interactions~\cite{ bib:shan-chen:1993, bib:shan-chen:1994}, finite-sized particles of arbitrary shape and wettability which can interact with the fluids as well as each other~\cite{ bib:ladd:1994:1, bib:ladd:1994:2, bib:jansen-harting:2011, bib:frijters-guenther-harting:2012, bib:guenther-frijters-harting:2014}, fully deformable particles~\cite{bib:krueger-frijters-guenther-kaoui-harting:2013,bib:kaoui-krueger-harting:2013,bib:krueger-kaoui-harting:2014}, and many other additives and additional interactions. Furthermore, the LB method is known for its relatively easy implementation and well-suitedness to massively parallel supercomputers~\cite{ bib:harting-harvey-chin-venturoli-coveney:2005, bib:groen-henrich-janoschek-coveney-harting:2011, bib:guenther-janoschek-frijters-harting:2013}.

\section{Domain sizes in particle-stabilised emulsions}

\begin{figure}[!htbf]
  \centering
  \includegraphics[width=0.45\textwidth]{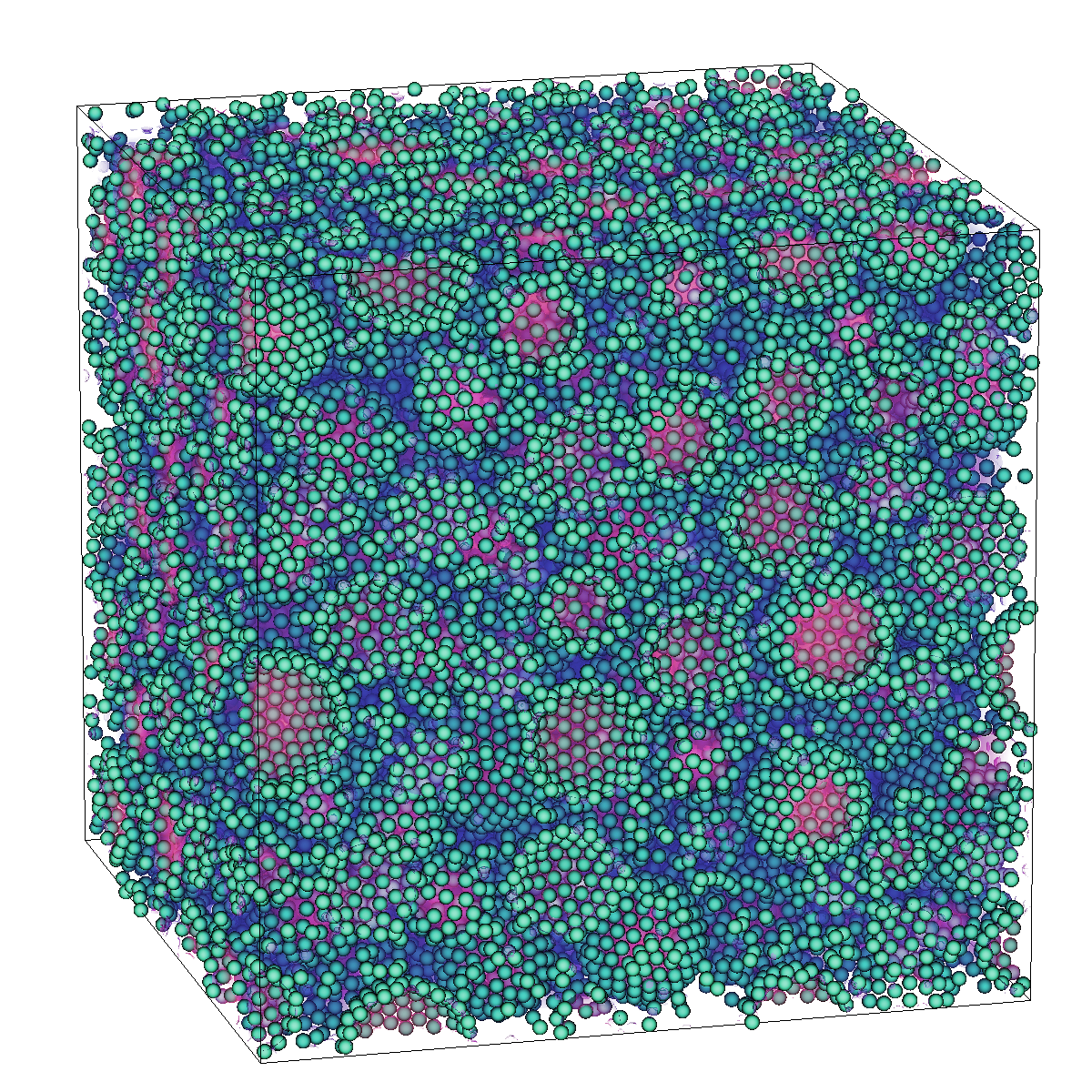}
  \includegraphics[width=0.45\textwidth]{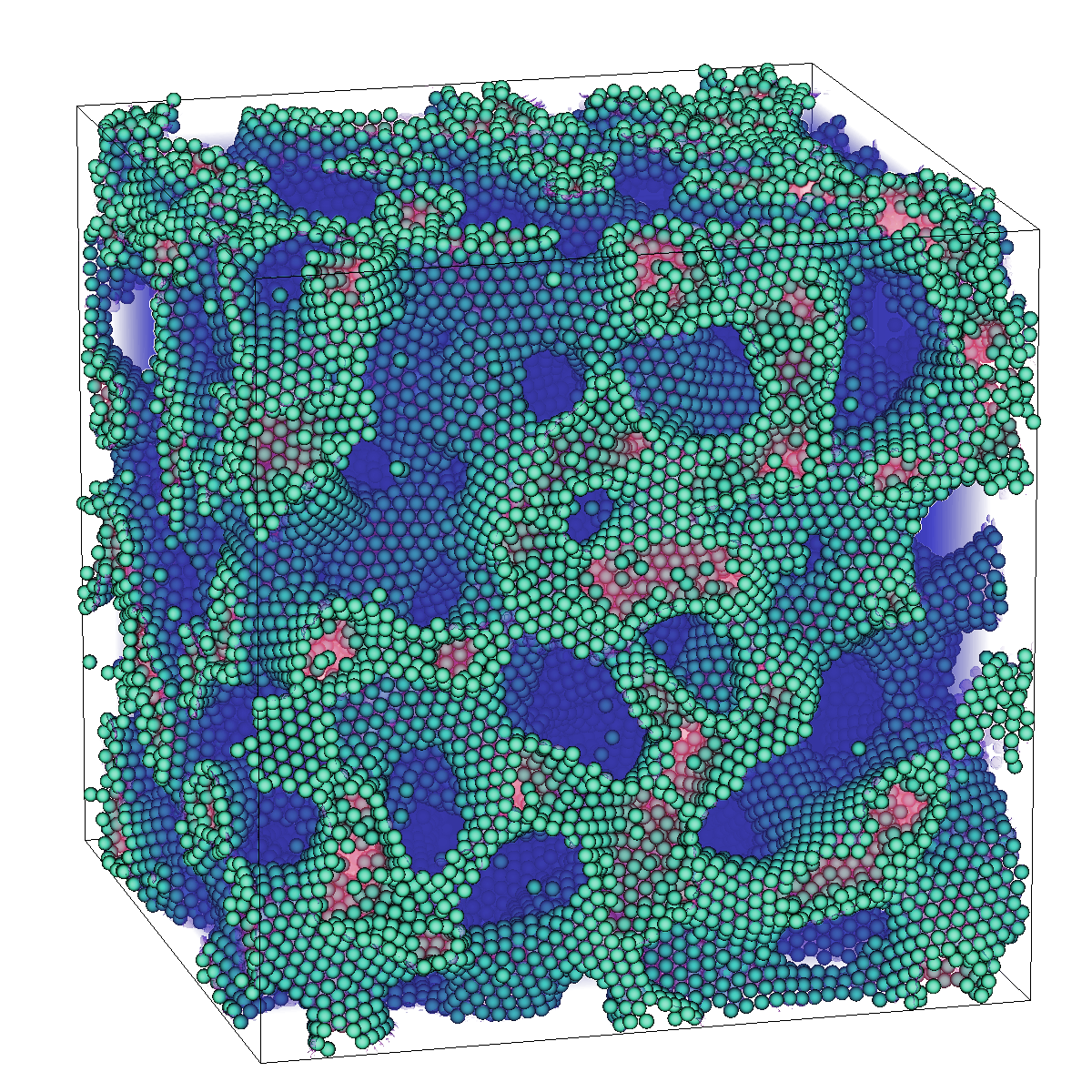}
  \caption{Examples of emulsion states stabilised by spherical particles with different chemical properties, which affects the preferred local curvature of the interfaces. Left: example of a Pickering emulsion --- discrete droplets (red) suspended in the medium (blue), and stabilised by particles (green). Right: example of a bijel --- two intertwined continuous domains.}
  \label{fig:emulsion-states}
\end{figure}

In recent decades, many branches of industry have started using nanoparticles to stabilise emulsions for various purposes (\emph{e.g.}~cosmetics, improved low-fat food products, drug delivery). Traditionally, amphiphilic surfactant molecules are often employed as emulsification agents, but their effects can be mimicked or supplemented by the use of nanoparticles. Particle-stabilised emulsions can be present in various forms, classified by the shapes and sizes of their contiguous fluid domains. The form that most resembles a traditional, surfactant-stabilised emulsion is the ``Pickering emulsion''~\cite{ bib:ramsden:1903, bib:pickering:1907}, which consists of particle-covered droplets of one fluid suspended in another fluid. A more recent discovery is the ``bijel'', or bicontinuous interfacially jammed emulsion gel, which is characterised by two large continuous intertwined fluid domains. This form was first predicted by numerical simulations~\cite{ bib:stratford-adhikari-pagonabarraga-desplat-cates:2005}, and shortly thereafter confirmed experimentally~\cite{ bib:herzig-white-schofield-poon-clegg:2007, bib:clegg-herzig-schofield-egelhaalf-horozov-binks-cates-poon:2007}. Examples of simulations of these systems are shown in Fig.~\ref{fig:emulsion-states}.

In this work we focus on the more traditional Pickering emulsions. We simulate the two fluids using the LB method, couple them using the Shan-Chen multicomponent model~\cite{ bib:shan-chen:1993}, and choose the strength of the interaction such that the fluids are immiscible. Particle-particle interactions are restricted to hard core repulsion and particle movement is treated \emph{via} molecular dynamics. The particles are coupled to the fluids following Ladd~\cite{ bib:ladd:1993}. The size of the stabilised droplets is often of interest (for example, it will strongly affect the mouthfeel of particle-stabilised food products) and to measure the size of the domains of such an emulsion one can use various techniques. Firstly, we define an order parameter $\phi(\vec{x})$ as being the difference in local density between the oil-like and water-like component in the emulsion, such that $\phi(\vec{x}) > 0$ for oil sites and $\phi(\vec{x}) < 0$ for water sites (a gradient exists near the diffuse interface). In Jansen et al.~\cite{ bib:jansen-harting:2011} and G{\"u}nther et al.~\cite{ bib:guenther-frijters-harting:2014} an average domain size for each of the three dimensions is calculated by considering the three-dimensional structure function based on the fluctuations of the order parameter $\phi$ in Fourier space. As an alternative or complementary method we can use the HK algorithm, where we choose $\varphi(\vec{x}) \equiv \phi(\vec{x}) > 0$ as the condition needed to decide whether a lattice site is part of a cluster, or part of the medium~\cite{bib:frijters-guenther-harting:2014b}. This corresponds to the sizes of the oil droplets being of interest and the water-like domain forming the medium. Once all sites have been labelled correctly, it is trivial to extract the size of each individual cluster by counting the number of sites with a particular label, and from this data the averages for the whole system are of course also easy to obtain. Statistics on the individual droplet sizes can be obtained in this way; this is not possible through the use of structure functions, which supply an average only. Furthermore, the Fourier-based method is negatively affected by the signal of the stabilising particles, which artificially decreases the magnitude of the length scales obtained in this way, while this problem does not exist for the HK-based method. However, the HK-based method does not supply information on the structure of bijels, as they contain only one cluster.

\begin{figure}[!htbf]
  \centering
  \includegraphics[scale=0.75]{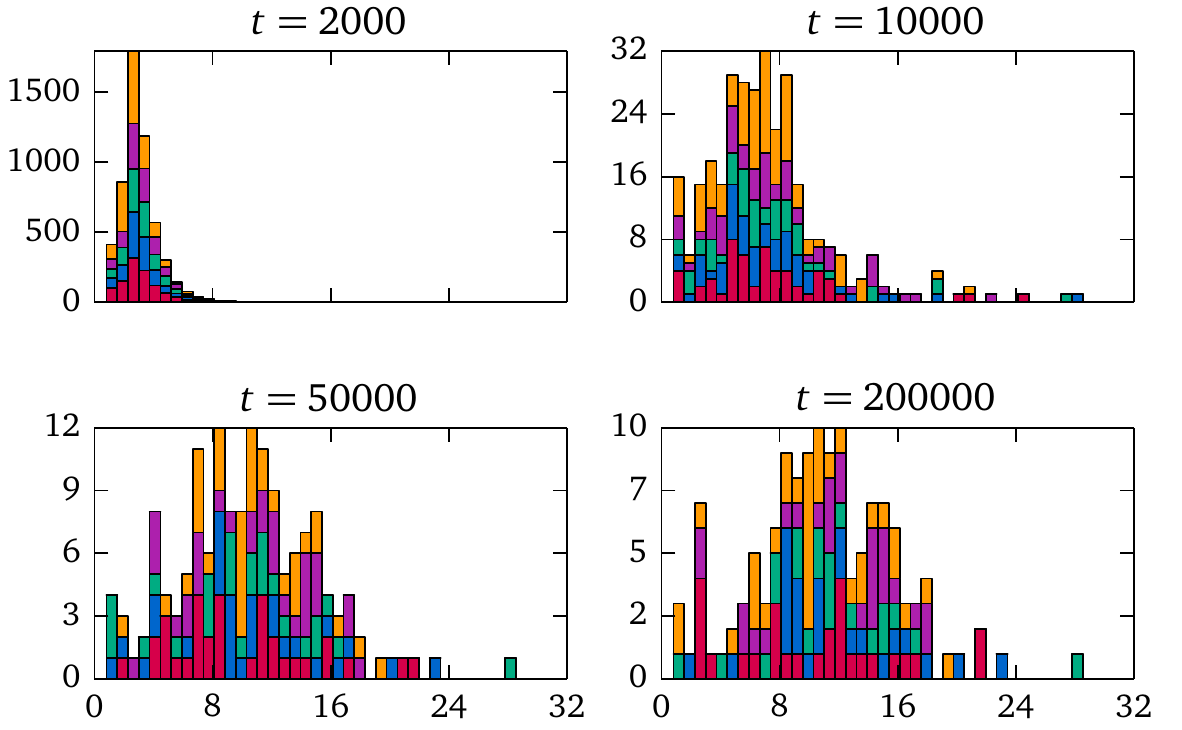}
  \caption{Histograms of the droplet size distributions of a Pickering emulsion as it evolves in time. The stacked bars of the histograms represent systems with identical physical properties but different random initializations. The radius $R_c$ of a spherical droplet with the volume of the corresponding cluster is used as a measure of the droplet size. Rapid nucleation takes place at the early stages of emulsion formation, resulting in the formation of many small droplets. This is followed by Ostwald ripening, which increases average droplet size and reduces the total number of droplets in the system. This process is blocked after some time, after which droplet growth only happens through coalescence of droplets and the evolution of the emulsion is almost halted.}
  \label{fig:droplet-histogram}
\end{figure}

We are interested in tracking the time evolution of the droplet sizes in a Pickering emulsion, starting from a random mixture of fluids and particles. The results of such a procedure can be seen in Fig.~\ref{fig:droplet-histogram}. The values shown are the sizes of the oil domains in units of radius of a spherical droplet with identical volume, and the stacked bars of the histograms denote systems with identical physical properties but different random initial distribution of the fluids and particles (this supplies better statistics). Three regimes have been identified in the time evolution of a Pickering emulsion~\cite{ bib:guenther-frijters-harting:2014}: after the simulation has been initialised, demixing quickly sets in due to the immiscibility of the fluids, causing the nucleation of droplets (which corresponds to the formation of many small domains). In this example, after $2000$ LB time steps the droplets have a narrow distribution centered around a radius $R_d \approx 3$. This radius is of the order of the interfacial thickness of the Shan-Chen multicomponent model and as such is not expected to be quantitatively accurate; however, this behaviour is qualitatively correct. Following this first period, the droplets continue to grow rapidly due to Ostwald ripening, which leads to both larger average droplet radius and a wider distribution as the system becomes less homogeneous. The total number of droplets shrinks rapidly. This process continues for some tens of thousands of time steps, after which Ostwald ripening is halted due to the almost maximal covering of fluid-fluid interfaces by particles. Then, the droplets can grow by coalescence only --- a rare phenomenon due to the particle layers --- and the average droplet radius stabilises at $R_d \approx 10$. It can be seen that the distribution of droplet sizes does not change significantly after $50000$ time steps. The availability of data on the distribution of the droplet sizes can help to, for example, identify parameters to help create monodisperse droplets of optimal size.

\section{Topology of deformable particles}

In general, the fluid properties in the interior of vesicles, capsules or biological cells differ from those outside. For example, the viscosity of the haemoglobin solution in the interior of red blood cells is about five times higher than the viscosity of the ambient blood plasma. Although the interior and exterior fluids are often assumed to be the same in simulations, researchers are showing an increasing interest in the effect of the interior fluid properties on the dynamics of a particle~\cite{bib:kaoui-krueger-harting:2013,bib:kaoui-jonk-harting:2014}.

In a transient situation, particles can deform and move, which makes it necessary to dynamically track regions inside and outside of a given closed surface area (\emph{e.g.}~the cell membrane). This can be done by using a ray-tracing algorithm \cite{macmeccan_simulating_2009} or solving a Poisson equation \cite{tryggvason_front-tracking_2001}; however, this is relatively expensive and not easy to parallelise. Instead, we have found that a variant of the HK algorithm is suitable to identify those regions.

\begin{figure}
  \centering
  \includegraphics[angle=90,width=0.38\textwidth]{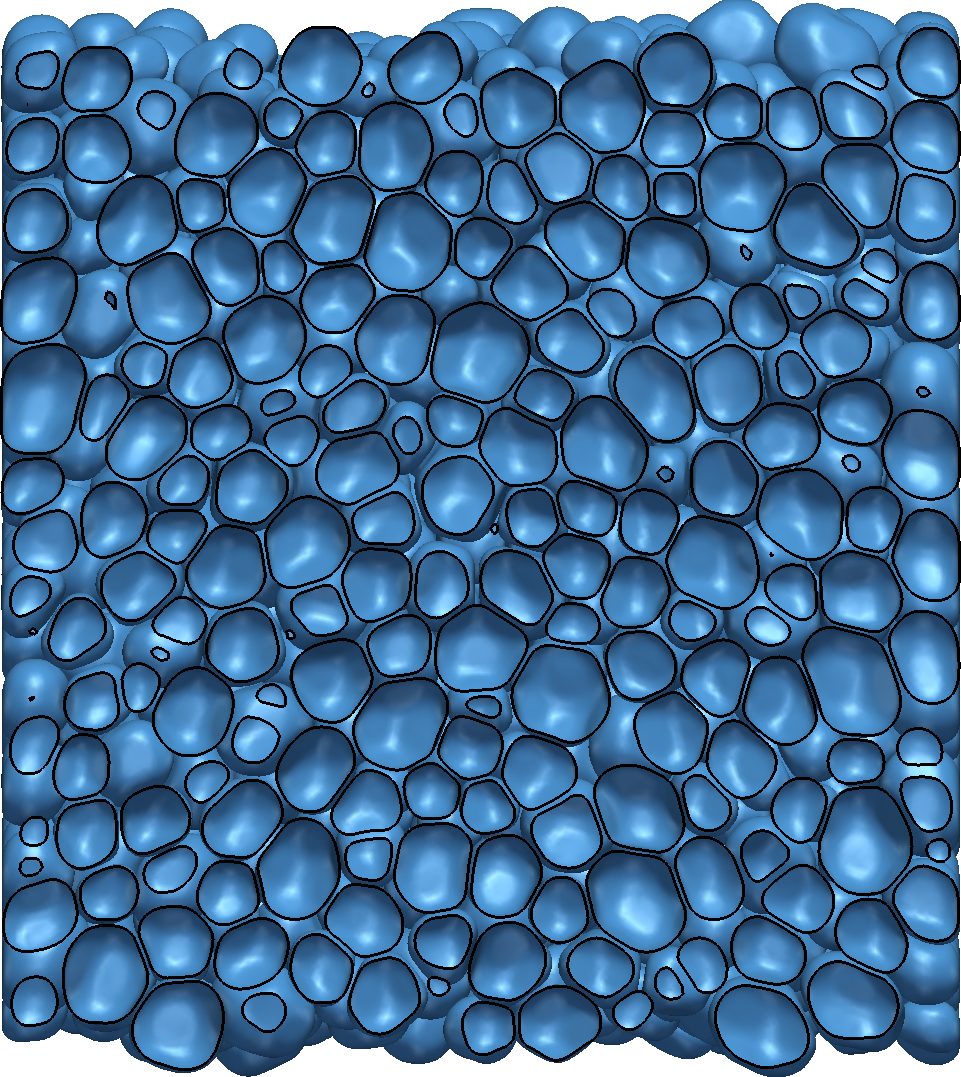} \qquad
  \includegraphics[angle=90,width=0.34\textwidth]{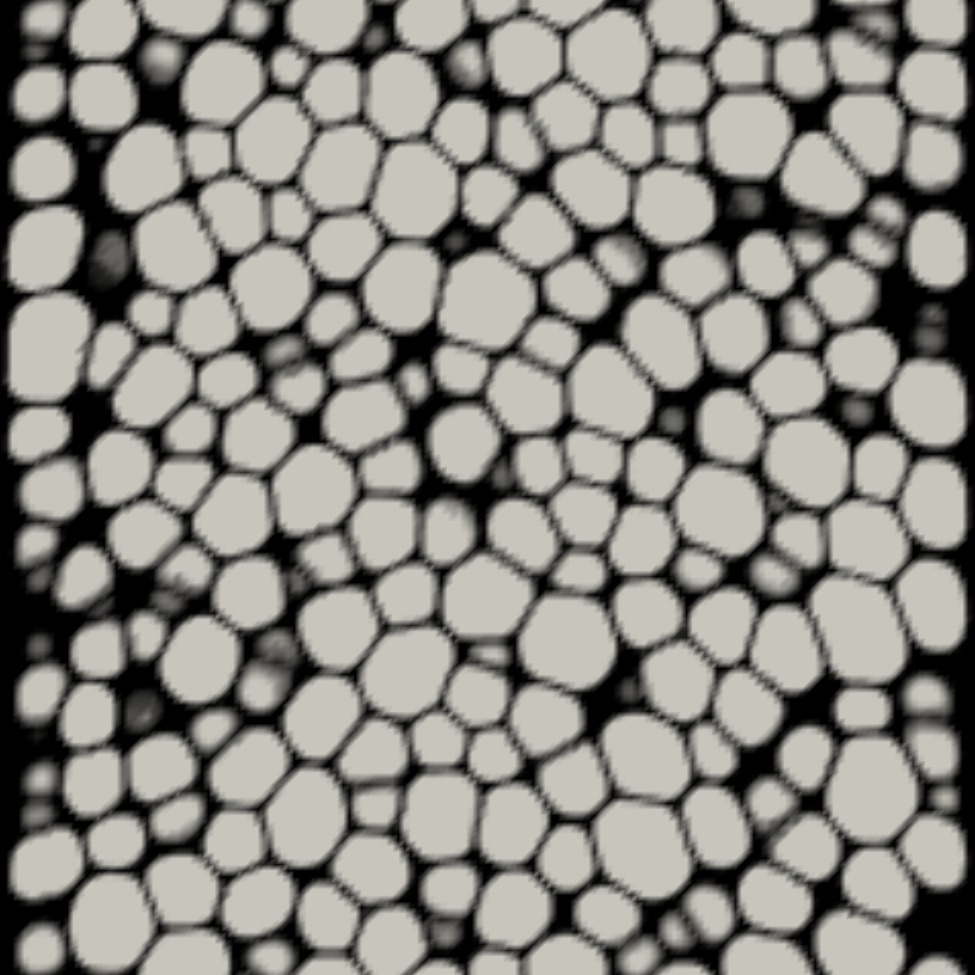}
  \caption{Snapshots of a dense suspension of deformable capsules. Left: visualisation of the membrane surfaces and a chosen cross-section. Right: value of the interior index field (black: $0$, outside; gray: $1$, inside) on the same cross-section. The HK algorithm has been used to obtain the index field on the lattice. A two-dimensional example has been shown for clarity --- the actual simulations are fully three-dimensional.}
  \label{fig:suspension_euler_lagrange}
\end{figure}

\begin{figure}
  \centering
  \begin{subfigure}[c]{0.475\textwidth}
    \centering
    \includegraphics[width=0.7\linewidth,trim=1cm 2cm 0.5cm 0.5cm, clip=true]{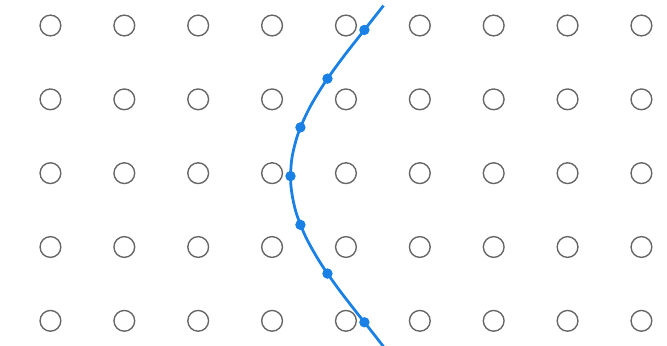}
    \caption{}
    \label{fig:deformable-0}
  \end{subfigure}
  \begin{subfigure}[c]{0.475\textwidth}
    \centering
    \includegraphics[width=0.7\linewidth,trim=1cm 2cm 0.5cm 0.5cm, clip=true]{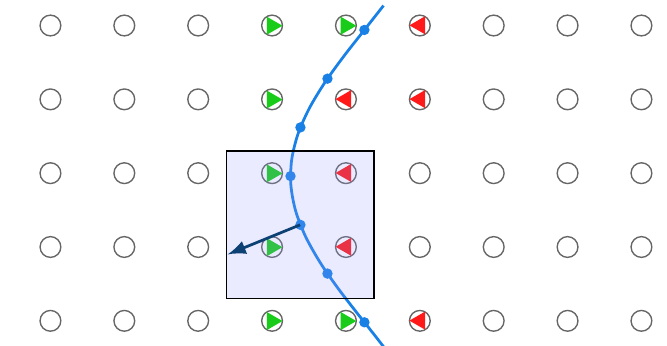}
    \caption{}
    \label{fig:deformable-2}
  \end{subfigure}
  \\[1ex]
  \begin{subfigure}[c]{0.475\textwidth}
    \centering
    \includegraphics[width=0.7\linewidth,trim=1cm 2cm 0.5cm 0.5cm, clip=true]{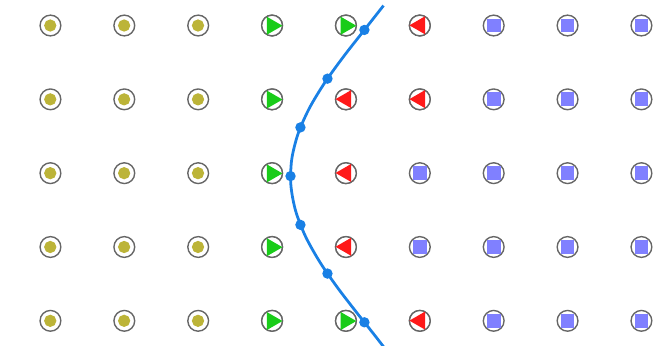}
    \caption{}
    \label{fig:deformable-3}
  \end{subfigure}
  \begin{subfigure}[c]{0.475\textwidth}
    \centering
    \includegraphics[width=0.7\linewidth,trim=1cm 2cm 0.5cm 0.5cm, clip=true]{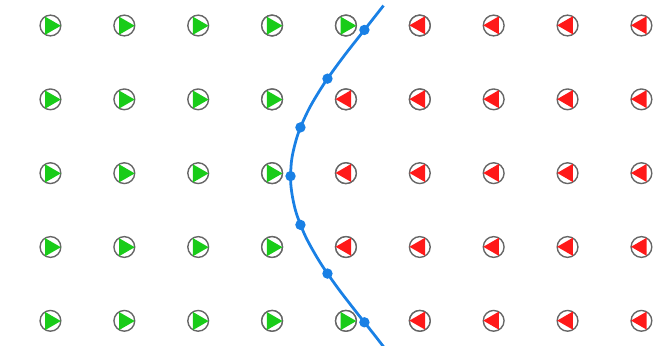}
    \caption{}
    \label{fig:deformable-4}
  \end{subfigure}
  \caption{Two-dimensional sketch of the application of the Hoshen-Kopelman algorithm to identify interior and exterior regions of particles. The initial state in (a) shows a lattice without any knowledge of the interior/exterior state of its sites (open circles) and a patch of an immersed particle surface. After identifying lattice sites close to the membrane (circles with triangles pointing to the right and left on different sides) by running over all membrane vertices and taking advantage of the known outward-facing normal vectors (b), the Hoshen-Kopelman algorithm is applied to all lattice sites which are not in direct vicinity of the membrane. As a result all clusters (circles with dots and squares) are identified (c). In the last step (d), those clusters are assigned to either the exterior or the interior region, depending on the nature of the surrounding near-membrane sites.}
  \label{fig:deformable}
\end{figure}

In our three-dimensional model, the closed surface of a deformable particle is described by a two-dimensional triangular unstructured mesh \cite{kruger_efficient_2011}. This mesh can move on top of the underlying lattice. Since the local fluid viscosity has to be known in order to perform the LB algorithm, each lattice site requires an additional up-to-date scalar value indicating its interior state. In our case, this interior state assumes $0$ if the site is sufficiently far outside and $1$ if it is sufficiently far inside. The remaining sites form a smooth transition region in direct neighbourhood of the membrane (with a thickness of usually one lattice constant $\Delta x$), where the interior index varies smoothly between $0$ and $1$ as illustrated in Fig.~\ref{fig:suspension_euler_lagrange}.

The algorithm to identify interior and exterior regions can be summarised as follows (see also Fig.~\ref{fig:deformable}):
\begin{enumerate}
 \item Introduce an auxiliary scalar index field $I$ and initialise it to a sufficiently negative value $I_0$ (with $I_0 < -0.5$) everywhere.
 \item Identify all lattice sites in direct vicinity of the membrane (Fig.~\ref{fig:deformable-2}). Run over all membrane mesh nodes and identify those lattice sites within a normal distance $d < 0.5$. By taking advantage of the known oriented membrane normal vectors, one can assign the value $d$ to those lattice sites outside and $-d$ to those inside (these values are all larger than $I_0$). This step runs in parallel. Afterwards, each lattice site has a proximity value $I$ either between $-0.5$ and $0.5$ (if close to a membrane) or it has retained the original value $I_0$ (if far away from the membrane).
 \item The next task is to identify clusters of all regions with the original index $I_0$. As additional condition we assume that interior and exterior regions are completely separated by a closed shell of sites with $-0.5 \leq I \leq 0.5$. This can be guaranteed by choosing a sufficiently dense membrane mesh.
 \item The HK algorithm is applied to those sites with $I = I_0$ and executed as described above. As a result, one finds a number of clusters which either describe pockets of interior or exterior fluid (Fig.~\ref{fig:deformable-3}).
 \item The last step is to decide which identified cluster is inside and which one is outside. We do this by assessing the value $I$ of neighbouring sites: an interior cluster is surrounded by sites with negative $I$ while an exterior cluster borders on sites with positive $I$ (Fig.~\ref{fig:deformable-4}). This concludes the identification algorithm.
\end{enumerate}

We note that the HK algorithm for the identification of interior and exterior regions does not have to be applied at every time step. Since the membranes move slowly on the lattice (much less than one lattice constant per time step), we take advantage of the fact that the interior index field is varying slowly and update only those lattice sites in the vicinity of the membranes. This algorithm is faster than the HK algorithm, but it still requires a valid index field at the previous time step. The HK algorithm is therefore used to initialise the interior index field or to refresh it from time to time.

\section{Conclusions}

The necessity to detect connected clusters in grid-based computer simulations
is a common problem as it was demonstrated by two examples in this article: we
reported on lattice Boltzmann simulations of droplets in an emulsion and the
flow of closed deformable particles in suspension. The analysis of the size
distribution of the droplets and the identification of interior and exterior
particle regions can be performed efficiently by using the Hoshen-Kopelman
algorithm with the correct conditions to define a cluster. An efficient
parallelisation of this algorithm is mandatory if it has to be applied on the
fly in massively parallel simulations. We demonstrated a possible
parallelisation scheme which follows the same cubic domain decomposition
strategy as generally used in lattice Boltzmann or other grid based
simulations. Using the same domain decomposition for the main simulation
algorithm and the Hoshen-Kopelman implementation is mandatory in order to avoid
unneccessary and costly data transfer between CPUs. By applying the algorithm
to an artificial test case, we demonstrated its good performance in a strong
scaling test performed on a BlueGene/Q supercomputer. Furthermore, we applied
the algorithm to the two examples from the field of soft matter physics as
mentioned above and demonstrated its usability for routine application in
massively parallel computer simulations. 

\section*{Acknowledgements}

Financial support is greatly acknowledged from NWO/STW (Vidi grant 10787 of J.~Harting) and FOM/Shell IPP (09iPOG14 - ``Detection and guidance of nanoparticles for enhanced oil recovery''). We thank the J{\"u}lich Supercomputing Centre and HLRS Stuttgart for the allocation of computation time.












\end{document}